\documentclass[manuscript]{acmart}

\AtBeginDocument{%
  \providecommand\BibTeX{{%
    \normalfont B\kern-0.5em{\scshape i\kern-0.25em b}\kern-0.8em\TeX}}}

\setcopyright{acmlicensed}
\copyrightyear{2026}
\acmYear{2026}
\acmDOI{XXXXXXX.XXXXXXX}

\acmISBN{978-1-4503-XXXX-X/18/06}

\usepackage{graphicx}
\usepackage[normalem]{ulem}
\useunder{\uline}{\ul}{}
\usepackage{framed}
\usepackage{tabularx}
\usepackage{multirow}
\usepackage{booktabs}
\usepackage[inline]{enumitem}
\usepackage{subcaption}

\usepackage{pgfplots}
\pgfplotsset{compat=1.12}
\usepgfplotslibrary{statistics}
\usepgfplotslibrary{colormaps}
\usepackage{tikz}
\usetikzlibrary{arrows, matrix, positioning, math, external, patterns}

\newcommand{\rebuttal}[1]{\textcolor{magenta}{#1}}

\begin{document}

\title{Co-Constructing Alignment: A Participatory Approach to Situate AI Values}

\author{Anne Arzberger}
\email{a.arzberger@tudelft.nl}
\orcid{0009-0003-7230-3975}
\affiliation{%
  \institution{Delft University of Technology (CS)}
  \city{Delft}
  \country{Netherlands}
}

\author{Enrico Liscio}
\email{e.liscio@tudelft.nl}
\affiliation{%
 \institution{Delft University of Technology (CS)}
 \city{Delft}
 \country{Netherlands} 
 }

\author{Íñigo de Troya}
\email{i.m.d.r.detroya@tudelft.nl}
\affiliation{%
 \institution{Delft University of Technology (TPM)}
 \city{Delft}
 \country{Netherlands} 
 }

\author{Maria Luce Lupetti}
\email{maria.lupetti@polito.it}
\affiliation{%
  \institution{Politecnico di Torino (Design)}
  \city{Torino}
  \country{Italy}}

\author{Jie Yang}
\email{j.yang-3@tudelft.nl}
\affiliation{%
  \institution{Delft University of Technology (CS)}
  \city{Delft}
  \country{Netherlands}}

\renewcommand{\shortauthors}{Arzberger, et al.}

\begin{abstract}
    As AI systems become embedded in everyday practice, value misalignment has emerged as a pressing concern. Yet, dominant alignment approaches remain model-centric, treating users as passive recipients of pre-specified values rather than as epistemic agents who encounter and respond to misalignment during interactions. Drawing on situated perspectives, we frame alignment as an interactional practice \textit{co-constructed} during human–AI interaction. We investigate how users understand and wish to contribute to this process through a participatory workshop that combines misalignment diaries with generative design activities. We surface how misalignments materialise in practice and how users envision acting on them, grounded in the context of researchers using Large Language Models as research assistants. Our findings show that misalignments are experienced less as abstract ethical violations than as unexpected responses, and task or social breakdowns. Participants articulated roles ranging from adjusting and interpreting model behaviour to deliberate non-engagement as an alignment strategy. We conclude with implications for designing systems that support alignment as an ongoing, situated, and shared practice.
\end{abstract}



\maketitle

\section{Introduction}

As Large Language Models (LLMs) become increasingly embedded in personal and professional practices, they mediate consequential decisions. In doing so, users frequently encounter model behaviour that feels inappropriate, misleading, or harmful \cite{tiwari2025biases}. This gap is commonly framed as a problem of alignment between AI system behaviour and human values \cite{gabriel2020artificial}. It has given rise to the field of \textit{AI value alignment}, which aims to ensure that AI acts in line with human values such as “helpfulness,” “harmlessness,” and “honesty” \cite{russell2019human, christian2020alignment, bai2022constitutional}.

Much of this work approaches values as abstract entities that can be specified and encoded prior to deployment. Dominant technical paradigms therefore seek to infer or optimise values during training, for instance, through reinforcement learning from human or AI feedback \cite{christiano2017deep, bai2022constitutional, shi2024wildfeedback, ouyang2022training}. Within these model-centric paradigms, users play at best a passive role, with values inferred indirectly from behavioural traces \cite{shi2024wildfeedback, gao2024aligning, wu2025aligning} or large-scale preference data \cite{christiano2017deep}.

In practice, however, human values are situated --shaped by context, lived experience, social identity, and local norms \cite{arzberger2024nothing}. Misalignment, therefore, does not simply reflect a failure to reproduce predefined values, but emerges when system behaviour clashes with users’ situated value commitments during interaction. Users are thus uniquely positioned to recognise such value misalignments as they arise in use \cite{jain2025human}, and they should have the agency to influence how systems respond when such values emerge in interaction \cite{polletta2019freedom}. From this perspective, alignment is not something achieved \textit{for} users, but something that must be continuously shaped \textit{with} them \cite{shen2024towards}.

We address this gap through the concept of \emph{co-construction of value alignment}. In this view, moments of misalignment are not failures to be corrected through model interventions, but sites where users recognise, interpret, and express what values matter and how they should be enacted. Rather than treating alignment as a one-time specification or inferring values from proxy signals \cite{bai2022constitutional, birhane2022power, delgado2023participatory, shi2024wildfeedback}, co-construction frames alignment as an ongoing process of interactional sense-making between users and models. Alignment, in this view, is not achieved once, but continuously enacted in use.

However, enabling meaningful user participation in value alignment is non-trivial. Values are inherently abstract and tacit \cite{rokeach1973nature}, making them difficult to externalise and articulate in interaction. While interactive and human-centred approaches to AI enable users to verify, correct, and steer model outputs during use \cite{raees2024explainable, belosevic2025user}, they tend to operate through more readily expressible proxies, such as general preferences (e.g., \cite{bo2025steerable, ma2024beyond, peng2025navigating}), which are distinct from values \cite{van2020values, zhi2024beyond}. This challenge is further compounded by predominantly text-based interaction paradigms (e.g., \cite{pei2023acquiring, fan2025user, kim2023cells, ma2025should, guo2026towards}), which constrain the expression of complex, contextual, and abstract concepts such as values \cite{jiang2023graphologue, qin2023chatgpt}. Recently emerging research on bidirectional, user-driven alignment more explicitly targets \textit{values} rather than related proxies, but either remains top-down and conceptual \cite{shen2024towards} or is confined to interactions within existing chatbot paradigms \cite{fan2025user}. As a result, the design space for engaging users in recognising, articulating, and negotiating values at run-time remains largely underexplored. Accordingly, we ask:

\begin{framed}
\textit{\textbf{RQ:} How do users of AI systems want to actively engage in the process of co-constructing situated value alignment?}
\end{framed}

To explore this design space, we developed a \emph{participatory workshop that facilitates the co-design of potential alignment practices}. Rather than limiting participation to existing alignment tools \cite{fan2025user} or prescribing alignment roles top-down \cite{shen2024towards}, the workshop invites participants to collectively imagine new forms of engagement with AI systems. It thus serves as an exploratory method to surface potential design spaces for supporting user participation in alignment.

The workshop unfolds in three phases: participants (0) document misaligned interactions using a Misalignment Diary; (1) unpack them by identifying situated values at risk; and (2) envision actions through metaphors, interaction strategies, and interface features that enable co-construction. We apply this workshop to the use case of LLMs as research assistants, which embed normative assumptions about research practices and style. We report on four workshops with 12 participants from diverse disciplines, analysed using reflexive thematic analysis.

Our findings show that value misalignments range from unmet task requirements to unmet social expectations. From there, participants identify both context-specific and personal values (e.g., research integrity and sustainability) at risk. Envisioned co-construction roles included actions of understanding and adjustment, supported by interface elements such as maps or sliders that repurpose and extend current system affordances. Importantly, the findings also indicate that the co-construction design space extends beyond current modes of interaction, encompassing actions such as non-engagement and building towards collective responsibility. Taken together, the workshop findings illustrate that co-construction constitutes a viable and underexplored design space for alignment.

By foregrounding user reflexivity and participation, our work reimagines alignment not as a one-time engineering target, but as an ongoing interactional process enacted by both models and users. We argue that responsible alignment efforts must treat users as agents, not merely sources of data or passive recipients, and attend to the situated co-construction of values at the moment of interaction.
\section{Related Work}
Research on value alignment examines how AI systems can be designed to act in accordance with human values \cite{russell2019human}. Core questions span both normative and technical challenges and include \textit{what} and \textit{whose values} should be reflected, and \textit{how} they can be captured and encoded. These concerns have motivated a range of technical approaches, including supervised fine-tuning \cite{ouyang2022training}, reinforcement learning from human feedback \cite{stiennon2020learning}, and direct preference optimisation \cite{rafailov2024direct}.

\subsection{Limitations of Model-Centric Value Approaches} 
Most contemporary alignment work adopts a model-centric paradigm in which values are treated as stable targets specified in advance and optimised at training time  \cite{arzberger2024nothing, shen2024towards}. Alignment is thus framed as a static system property, evaluated by how closely model outputs conform to predefined values \cite{Liscio2022a, sorensen2024value, Hendrycks2021}. Values are typically defined before deployment, borrowing from established psychological \cite{Schwartz-2012-ORPC-BasicValues} or evolutionary \cite{Graham2009} theories, or framed as variants of ``helpfulness,'' ``harmlessness,'' or ``honesty'' \cite{dahlgren2025helpful, christian2020alignment, bai2022constitutional}. Usually, they are operationalised using large-scale preference datasets, often derived from crowdworker judgments and aggregated into singular reward signals \cite{delgado2023participatory, christiano2017deep, qiu2022valuenet, Hoover2020}. Within this paradigm, users contribute data, while having limited influence over how values are defined, or interpreted \cite{delgado2023participatory}.

This view of alignment presumes that relevant values can be defined independently of the contexts in which AI systems are used, collapsing complex ethical judgments into decontextualised data points or aggregate preferences \cite{arzberger2024nothing}. This optimisation logic flattens disagreement, contestation, or negotiation around values into a single “ground truth” signal \cite{aroyo2015truth}. Model-centric strategies also undervalue the epistemic role of users, limiting them to passive recipients rather than active interpreters or negotiators of model behaviour \cite{delgado2023participatory}. 

These simplifications have concrete downstream consequences. A model trained to optimise a single aggregated preference signal struggles to navigate the nuances of particular domains, communities, or interpersonal situations. Such models lack access to emergent, tacit, or domain-specific knowledge and cannot recognise when behaviour that is globally reasonable becomes locally inappropriate or harmful \cite{gabriel2020artificial, bommasani2021foundation}. As users and contexts diverge from what is represented in training data, system behaviour becomes increasingly brittle, producing misalignments that disproportionately affect already marginalised communities \cite{wong2020cultural}.

In this work, we break with the notion of values as readily codifiable predictors of ``good'' behaviour. Instead, we draw on work that conceptualises values as \emph{situated}, shaped by social context, lived experience, and interactions \cite{arzberger2024nothing, liscio2021axies, pommeranz2012elicitation, deWet2018}. From this perspective, misalignment emerges in use when models fail to accommodate evolving goals or contextual expectations. Alignment, in turn, becomes an ongoing interactional practice, requiring systems that support users in articulating, negotiating, and contesting values as they arise, rather than presuming they can be fixed in advance.

\subsection{Toward Participatory Alignment}
In this work, we focus on user participation during interaction, i.e., how users can shape AI system behaviour during use, as we are interested in how users participate in the alignment of system behaviour as it unfolds in use. This differs from approaches that involve users in informing system design during development, e.g. through interviews or design workshops (e.g. \cite{bhargava2019gobo, krishnaswamy2017participatory, chandrasekharan2019crossmod, scott2022algorithmic, young2019toward, holten2020shifting}).

Interactive and human-centred approaches increasingly enable users to verify, correct, and steer outputs during interaction with AI systems \cite{raees2024explainable, belosevic2025user}. Users can, for instance, support factual verification and error correction \cite{chen2024your, laban2024beyond, lee2025veriplan}, maintain human oversight in expert workflows \cite{gupta2025autosumm, dai2023llm}, engage in co-creative or assistive tasks such as writing or planning \cite{gero2022sparks, kim2023metaphorian, yang2022ai}, or personalise model behaviour to individual preferences \cite{peng2025navigating, bo2025steerable}. A range of interface paradigms has emerged to facilitate such engagement, including dialogue-based prompting \cite{pei2023acquiring, chen2024your, oppenlaender2025prompting, peng2025navigating, suh2024luminate, theophilou2023learning}, structured elicitation (e.g., questionnaires) \cite{chun2025planglow, bo2025steerable}, decomposition and clarification interfaces using graphical representations \cite{jiang2023graphologue, ma2024beyond, kim2023cells}, and control mechanisms such as sliders for navigating behavioural dimensions \cite{bo2025steerable, viegas2023system, chen2024designing}.

Much of this work is situated within specific domains or workflows \cite{belosevic2025user}, such as education \cite{theophilou2023learning}, or content generation \cite{gero2022sparks}, where user involvement is structured around task-specific goals. Such approaches do not establish a general interaction paradigm for addressing misalignments across contexts. More fundamentally, these systems are largely motivated by improving task performance, trust, usability, and controllability \cite{raees2024explainable, belosevic2025user}, and therefore operationalise user participation through proxies such as preferences (e.g. \cite{bo2025steerable}). While effective for adapting system behaviour, these constructs are conceptually distinct from values \cite{van2020values} and have been questioned as adequate representations of human values in alignment contexts \cite{zhi2024beyond}. This limitation is further reinforced by the predominance of text-based interaction paradigms (e.g. \cite{pei2023acquiring, fan2025user, kim2023cells, ma2025should, guo2026towards}). While flexible, text constrains the expression of abstract, structured, or relational concepts \cite{jiang2023graphologue, qin2023chatgpt}, and often results in verbose and effortful exchanges \cite{chang2002effect}. These challenges are particularly pronounced for values, which are abstract, tacit and challenging to articulate \cite{rokeach1973nature}. Consequently, interaction tends to gravitate toward more readily expressible proxies, leaving value considerations implicit: correcting or steering outputs may reflect concerns such as fairness or appropriateness, but current systems rarely externalise or support reasoning about these concerns explicitly. Relevant work focusing specifically on explicit user participation in LLM value alignment at run-time remains limited, and either focuses on interaction dynamics within standard chatbot affordances \cite{fan2025user}, or remains largely conceptual and top-down \cite{shen2024towards}. Finally, many of these approaches originate in personalisation paradigms and therefore focus on enabling interaction with the \textit{individual} (e.g. \cite{tankelevitch2024metacognitive, ma2024beyond, gershoff2008s}), limiting the design space to one-to-one interactions and overlooking collective forms of participation. 

We aim to translate these established commitments to user participation into the context of LLM value alignment, conceptualising it as a process of \textit{co-construction} in which users actively engage in interpreting and negotiating what values matter and how they should be enacted in context. This shifts the focus from operating on outputs to engaging with the interpretations and values that shape them. To explore this design space beyond prompting and one-to-one interactions, we introduce a participatory workshop that enables users to reflect on lived experiences of misalignment, articulate emerging values at risk, and imagine new forms of interaction that support value alignment at run-time.
\section{Research Methodology} 
We adopted a participatory methodology anchored in generative design research (GDR) and Interaction Vision (IV) methods to surface and reflect on situated value misalignments as they arise in everyday AI use, and to explore how users envision acting on these misalignments through interaction and interface design. We empirically examined this approach through a series of workshops focused on the use of LLMs as research assistants.

\subsection{Workshop Theory: Generative Design Research and Interaction Vision}
The workshop design draws on participatory and GDR traditions in HCI, which treat users as experts in their own practices, values, and alignment needs \cite{delgado2023participatory, sanders2012convivial}. A central challenge in engaging users around values is that values are often tacit, guiding action without being easily articulated \cite{friedman2013value, sadek2024designing}, while participants’ ability to imagine alternative AI interactions is constrained by existing systems and norms \cite{sanchez2025let, pasman2011interaction}.

To address these challenges, we drew on GDR methods \cite{sanders2012convivial}, which surface tacit knowledge through making, doing, and reflection rather than verbal articulation alone. We operationalised this approach through a GDR workbook in the form of a \textit{misalignment diary}, which sensitised participants to moments of misalignment as they occurred in their everyday use of LLMs as research assistants. 
During the workshops, taking inspiration from IV \cite{pasman2011interaction} and the New Metaphor Toolkit \cite{ricketts2019mental}, participants engaged in generative exercises supported by (1) a curated image set spanning abstract to concrete metaphors for co-construction actions, and (2) a collection of interface elements (e.g., buttons, sliders, visualisations, and alternative interaction forms). Both sets can be found in Appendix \ref{appendix:C}. These materials served as prompts to help participants externalise situated values, imagine alignment practices beyond present constraints and intuitive interventions, and envision interfaces that support co-construction in use.

\subsection{Workshop Setup}

\textbf{Use Case: LLMs as Research Assistants.}
We situated the workshops around LLMs used as research assistants, which increasingly mediate activities such as literature review, drafting, and idea generation (e.g., \cite{schmidgall2025agent, whitfield2023elicit, sohail2025using}). These systems embed normative assumptions about evidence, style, and epistemic practice, often privileging dominant academic traditions or concise, persuasive outputs, yet such defaults are typically fixed at design time and difficult to contest in use \rebuttal{\cite{Bender_Gebru_McMillan-Major_Shmitchell_2021, Weidinger_Mellor_2021}}. Research practices are deeply contextual and value-laden, shaped by disciplinary norms, epistemic commitments, and concerns such as inclusivity and sustainability. This makes research assistance a particularly generative site for studying situated alignment. By focusing on this use case, we examine how diverse publics, with varying familiarity with research and LLMs, can participate meaningfully in shaping model behaviour at runtime.

\textbf{Participant Recruitment.} 
We recruited 12 participants through personal academic networks to achieve diversity across disciplines, including engineering, social sciences, and the humanities, with varying notions of rigour, evidence, and objectivity. Participants also varied in their experience using LLMs for research, ranging from novice to expert users. To support shared vocabulary and discussion, participants with similar disciplinary backgrounds were grouped within workshops, while the overall sample enabled comparison across epistemic traditions. Demographic details and group divisions are provided in Appendix \ref{appendix:B}.

\textbf{Format and Data Collection.} 
We conducted four workshops (one in person and three online), each lasting roughly three hours. This hybrid format enabled participation across geographic and disciplinary contexts, though we acknowledge trade-offs in engagement and immersion (Section \ref{sec:limitations}). All sessions were audio-recorded and transcribed. We additionally collected written responses, misalignment diaries, screen recordings, and photographs of the generated artefacts. The study received institutional ethics approval from \anon[an institutional ethics review committee]{(TU Delft, approval code 187487)}.

\subsection{Workshop Analysis}
We analysed workshop materials using reflexive thematic analysis (RTA), inspired by Braun and Clarke \cite{braun2006using, clarke2017thematic}. Our primary analytic focus was on participant-generated artefacts, including misalignment diaries, written reflections, and design outputs, which were explicitly created to capture situated understandings of (mis)alignment practices. Analysis proceeded through repeated familiarisation with the data, followed by inductive coding grounded in participants’ language and representations. We coded data and iteratively refined themes on the digital brainstorming platform Miro\footnote{\url{www.miro.com}}. Codes captured misalignment experiences, values at stake, intuitive alignment actions, metaphors of co-construction, and proposed interface elements. We organised analysis around the workshop exercises, enabling comparison within and across groups. Codes were iteratively refined and clustered into higher-level themes through recursive movement between data and interpretation (see conceptual coding model in Appendix \ref{appendix:d}). Workshop transcripts were not systematically coded; instead, they were used selectively to contextualise artefacts, clarify ambiguities, and enrich interpretation where workshop materials fell short. This analytic choice reflects the workshop’s participatory design emphasis on participant-generated artefacts as the primary locus of reflection, rather than on conversational interaction alone.

\anon[Two researchers]{Anne and Enrico} independently clustered the dataset before jointly discussing interpretations, tensions, and positionalities \cite{hopf1993verhaltnis, braun2019reflecting}. Our goal was not to establish coding reliability, but to foreground plurality and reflexivity in interpreting inherently situated and contested value judgments. In line with RTA’s epistemological commitments \cite{braun2006using, braun2019reflecting}, we did not calculate inter-coder reliability; instead, consensus-building functioned as an interpretive practice rather than a metric of objectivity \cite{mcdonald2019reliability}.
\section{Workshop Design}
\begin{figure}[ht]
    \centering
    \includegraphics[width=1\linewidth]{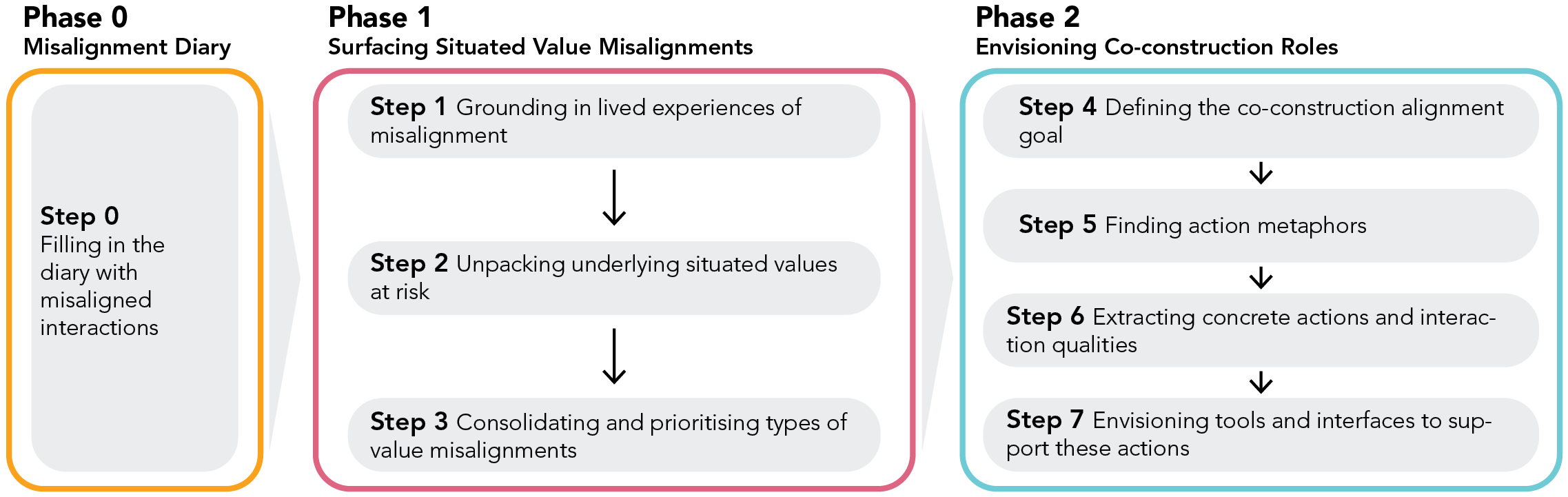}
    \caption{Overview of the three workshop phases. Phase 0 uses a diary to sensitise participants to different forms of misalignment in AI interactions. Phase 1 foregrounds the negotiation of situated values shaping misaligned AI interactions. Phase 2 moves from problem articulation to solution-making through discussion of actions and interface interventions toward a self-defined alignment goal.}
    \Description{Three-phase participatory workshop process for envisioning user co-construction of AI value alignment. The figure shows a linear sequence of three phases arranged from left to right. Phase 0 introduces a misalignment diary activity that sensitises participants to different misalignments in their everyday interactions with artificial intelligence systems. Phase 1 focuses on group reflection, where participants unpack and negotiate the situated values underlying these misaligned interactions. Phase 2 shifts to solution-making, depicting discussion of concrete actions and interface ideas to work toward a self-defined alignment goal.}
    \label{fig:workshop_phases}
\end{figure}

Our workshop supports users of AI systems in capturing use-specific experiences of misalignments, unpacking and negotiating the situated values at risk, and envisioning user interventions beyond intuitive prompt engineering. Fig.~\ref{fig:workshop_phases} presents an overview of the workshop, composed of three phases: Phase 0 is an asynchronous stage during which participants are asked to record their experiences of misalignment. Phases 1 and 2 constitute a facilitated small-group workshop where participants surface value misalignments and envision co-constructing roles, respectively.

\subsection{Phase 0 - Misalignment Diary}
Phase 0 grounds the co-construction process in participants’ everyday AI use through a generative design probe \cite{sanders2012convivial}: the \emph{Misalignment Diary}. The diary supports recording and reflection on misalignments as they occur, defined broadly as: \textit{moments where the AI’s response didn’t quite fit the situation, when what it did or assumed didn’t match what made sense, felt right, or seemed important in context.} Two weeks before the following workshop phases, we distributed a digital form asking participants to record instances of misalignment they encountered in their use of LLMs. Entries may include screenshots, notes, or reflections. Precisely, participants are asked to document:
\begin{enumerate*}
    \item \textit{Initial Prompting} (what they were trying to achieve and why);
    \item \textit{Misaligned Response} (where the system output diverged from expectations);
    \item \textit{Intuitive Intervention} (any immediate action taken to address it); and
    \item \textit{Final Outcome} (whether interventions were successful and final outcomes aligned/misaligned).
\end{enumerate*}
Fig.~\ref{fig:examplephase0} displays the example of one of Participant 7 (P7) diary entries.

\subsection{Phase 1 - Surfacing Situated Value Misalignments} 

Phase 1 builds on the diary material to move from \textit{what} happened to \textit{why} it mattered, surfacing the \textit{situated values} underlying users’ experiences. The purpose of this process is to uncover the deeper stakes, ensuring that solutions proposed in Phase 2 target the root causes rather than the surface-level problems.

\begin{figure}[h!]
    \centering
    \includegraphics[width=1\linewidth]{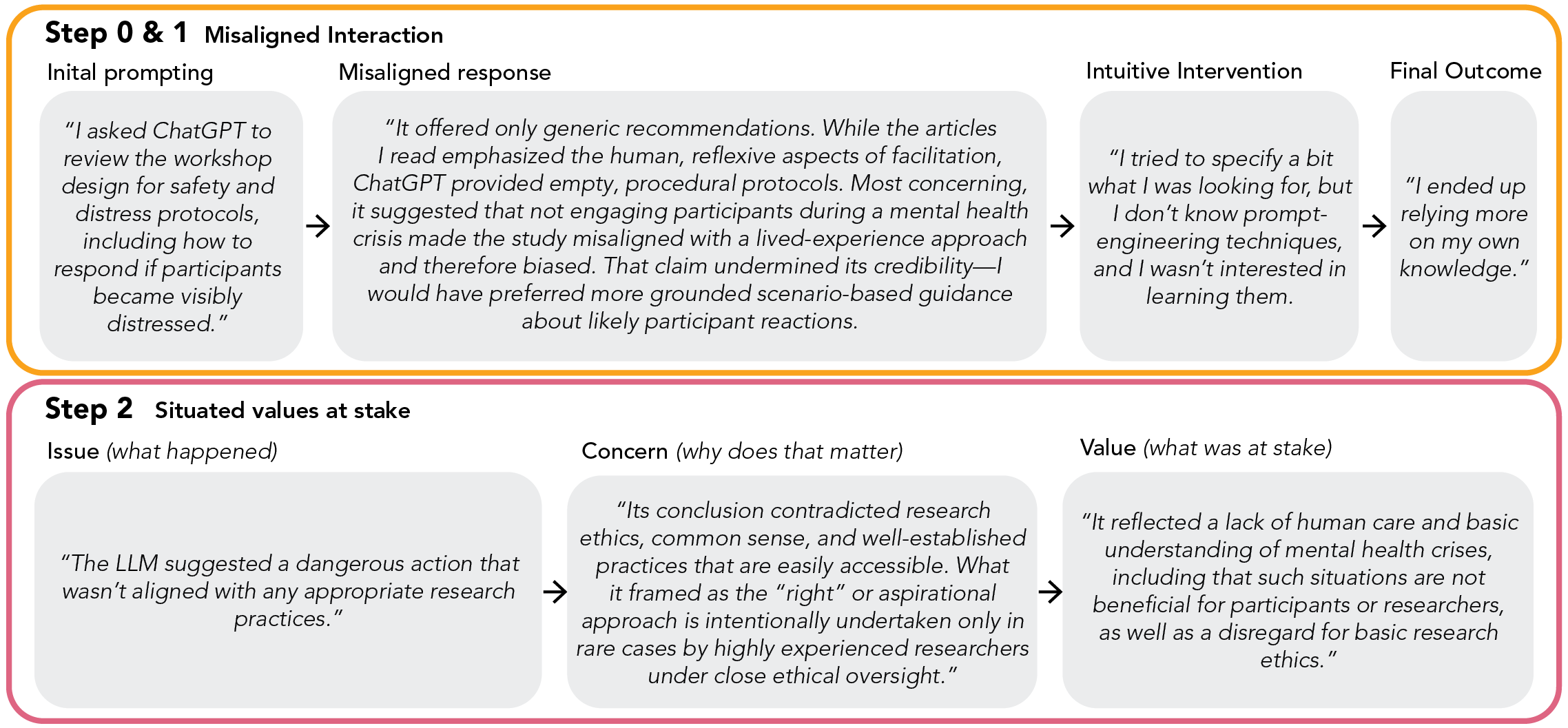}
    \caption{Example of a misaligned interaction traced from initial diary entry to concern and values at stake across Phases 0–2. The example from P7 documents an initial prompt, a misaligned response, an intuitive intervention, and the final outcome when using ChatGPT to assess workshop safety, followed by a reflection on why the misalignment mattered and values were implicated.}
    \Description{Tracing a misaligned interaction from diary entry to concern and values at stake.
The figure shows how a single misaligned interaction is documented and progressively elaborated across early workshop phases. A diary entry captures an initial prompt, a misaligned response, an intuitive intervention, and the final outcome, illustrating how the participant received workshop safety suggestions they perceived as unethical and relied on personal judgment rather than prompt refinement. The interaction is then reframed as a concern, identifying risks to human care, shared understandings of mental health, and research ethics.}
    \label{fig:examplephase0}
\end{figure}

\paragraph{Step 1: Grounding}
Each participant shares two diary entries, briefly describing what happened and why it mattered. These shared examples serve as the basis for a collective understanding and analysis.

\paragraph{Step 2: Zooming out}
Participants individually reflect on what their misalignments reveal about AI behaviour, then reconvene to discuss multiple interpretations. Step 2 guides them to move from concrete experience (\textit{what happened?}) to concerns (\textit{why did it matter?}) to situated values (\textit{what was at stake?}). They are given a definition of values as: \textit{principles, commitments, and qualities that matter in your work and that you want your interactions with AI to respect and support}, together with examples of what values entail (e.g., how you want to think ``critically, creatively'') and do not entail (e.g, personal taste ``I like blue better than brown''). Participants are encouraged to stay connected to context rather than drifting toward high-level value terms (e.g., fairness, autonomy). We deliberately do not provide predefined value lists to prevent priming and allow contextual specificity.
Fig.~\ref{fig:examplephase0} shows the example of step 2 for P7.

\paragraph{Step 3: Consolidating}
To further abstract from the specific individual experiences and to identify common patterns of value misalignments within the context of use, the group next needs to discuss and negotiate to form clusters. They are free to choose clusters based on values found at stake, or different mechanisms they see putting these values at risk. Ultimately, the group is asked to prioritise a single form of misalignment to address in the second half of the workshop.

\subsection{Phase 2 - Envisioning Co-construction Roles}
Phase 2 invites participants to reimagine alignment as a configuration of roles, relationships, and interaction qualities between users and AI systems. Addressing the tendency to default to prompt refinement or system fixes \cite{fan2025user}, we draw on generative design traditions \cite{sanders2012convivial} to decouple imagination from technical feasibility, allowing participants to articulate alignment as relational dynamics before translating them into actionable features.

\begin{figure}[ht!]
    \centering
    \includegraphics[width=1\linewidth]{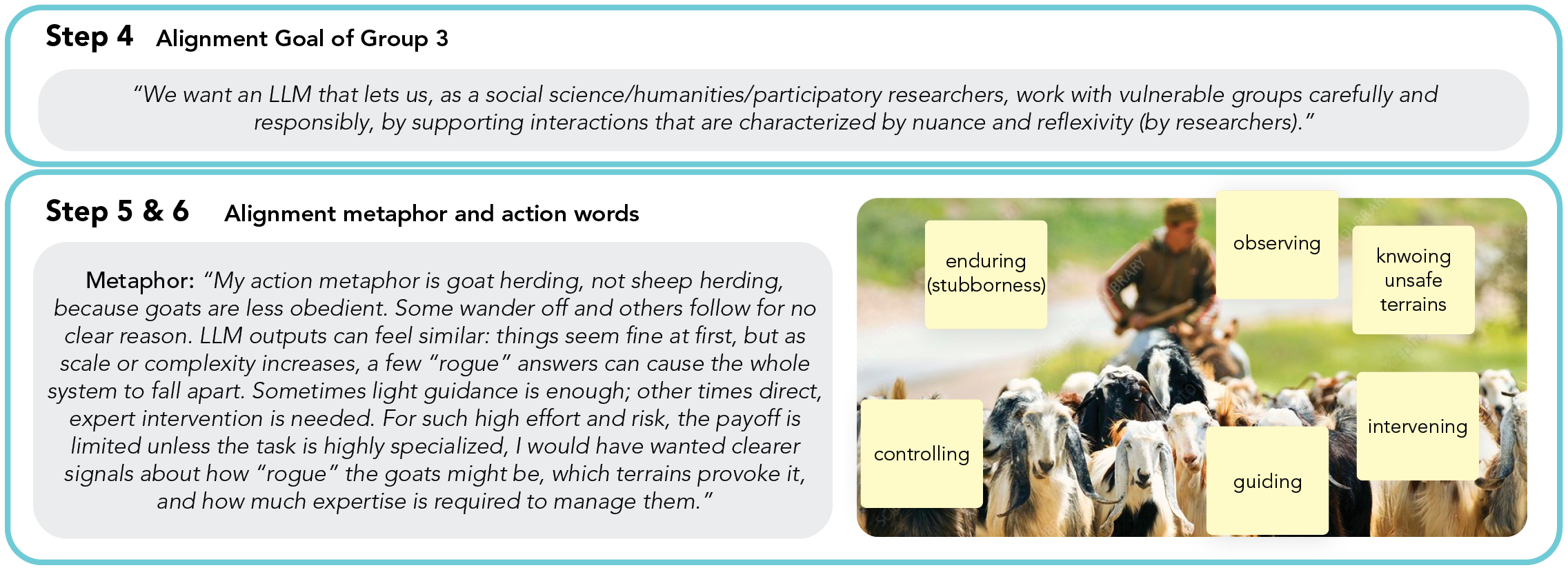}
    \caption{From a shared alignment goal to an action metaphor in Steps 4–6. P7 and Group 3 define an alignment goal emphasising nuance and reflexivity, which P7 further translates into user actions through a goat-herding metaphor.}
    \Description{Defining a shared alignment goal and translating it into reflexive user actions. The figure shows how a prioritised misalignment is translated into a shared alignment goal by workshop group 3, highlighting nuance and reflexivity as desired qualities of aligned interactions. It then illustrates how participant 7 extends this goal using a goat-herding metaphor to express how reflexivity and nuance can be enacted through user actions, including enduring and intervening.}
    \label{fig:step456}
\end{figure}

\paragraph{Step 4: Defining Alignment Goal}
Participants restate their chosen form of misalignment as an aligned state, focusing on the user, not the AI action: “We want an LLM that lets us, as a \_\_\_ researchers, \_\_\_, by supporting interactions that are characterised by \_\_\_ .” For example, Group 3 (P7's group) filled in the sentence as displayed in Fig.~\ref{fig:step456}.

\paragraph{Step 5: Finding Action Metaphors}
Participants individually select 2–3 images that capture the felt dynamics of the aligned state. Using metaphors and images helps participants move beyond fixation on existing interaction possibilities and express tacit relational understandings that are difficult to verbalise \cite{pasman2011interaction, ricketts2019mental}. This intuitive entry point also levels disciplinary differences, enabling participants with diverse backgrounds to articulate shared aspirations without technical vocabulary. We provide a curated image set, ranging from concrete actions to abstract representations (Appendix \ref{appendix:C}), pre-labelled following the New Metaphor Toolkit\footnote{\url{https://imaginari.es/new-metaphors/}}. Fig.~\ref{fig:step456} shows one metaphor selected by P7.

\paragraph{Step 6: Extracting Concrete Interaction Qualities}
As a group, participants then present the metaphors and annotate them to extract descriptive adjectives, verbs, and relational cues from the images (e.g., guiding, questioning, dialogical, transparent) to define \textit{action words} that extend our pre-assigned labels. These qualities act as bridges between affective imagination and designable interaction properties.
Fig.~\ref{fig:step456} shows the annotations of an image chosen by P7.

\begin{figure}[h!]
    \centering
    \includegraphics[width=1\linewidth]{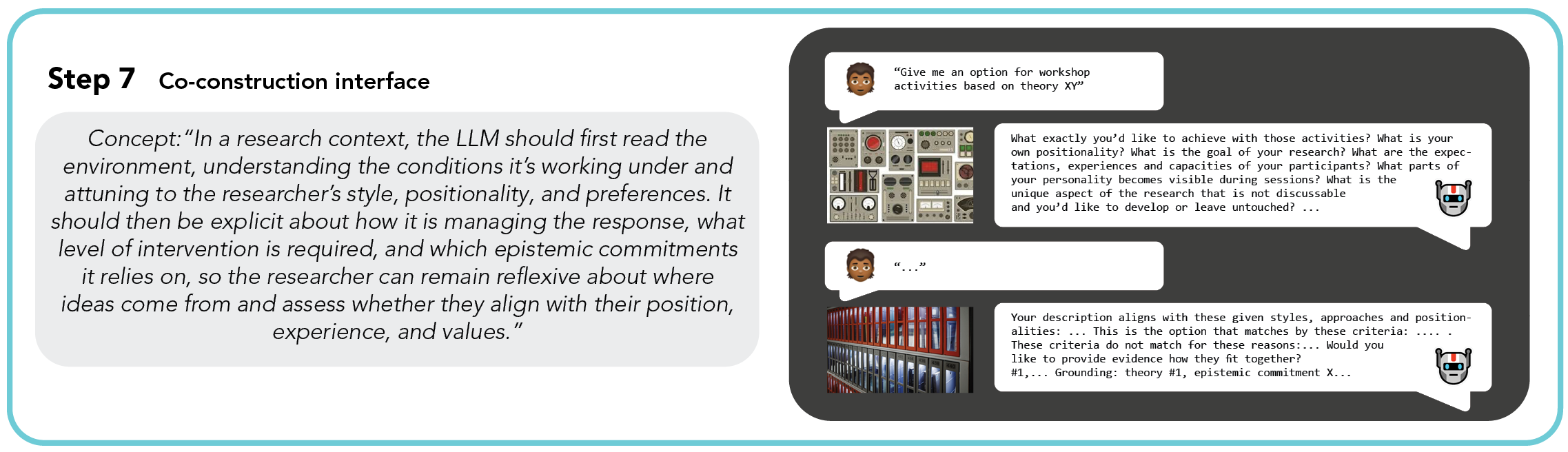}
    \caption{P7-envisioned interface for Step 7, supporting reflexive alignment through visible model positionality.}
    \Description{Interface concept for examining and responding to model positionality. The figure illustrates the conceptual interface of participant 7, where a language model explicitly presents its positional stance to the user. This visibility facilitates a reflexive comparison between the model’s stance and the researcher’s own positionality, enabling more deliberate adjustments of prompts and interactions.}
    \label{fig:step7}
\end{figure}

\paragraph{Step 7: Tools and Interfaces}
Participants individually translate envisioned actions into tools, interface features, or feedback mechanisms. Facilitators prompt them to remain grounded in the values, alignment goals, and annotated action words, using metaphors as starting points for ideation. Participants work with an empty interface canvas and a set of digital and physical control elements, such as buttons, sliders, and monitors (see Appendix \ref{appendix:C}). This synthesis connects imaginative articulation with practical grounding, bridging expressed values and enactable relationships and generating insights for interaction design and broader alignment governance. Fig.~\ref{fig:step7} shows an example from P7.

\section{Results}
We present the insights into situated misalignment and user actions discussed and envisioned during our workshop.

\subsection{Theme 1: Unpacking Situated Value Misalignment}
We describe the common trends participants articulated during Phase 1, focusing on 
\begin{enumerate*}
    \item the misalignment experiences that the participants reported in their diaries,
    \item the intuitive intervention participants took to address the misalignments, and
    \item the situated values identified to be at risk in those experiences.
\end{enumerate*}

\subsubsection{What misalignments were experienced in LLM use}
To understand how misalignment materialises in situated use, participants recorded instances of misaligned interactions in a misalignment diary. Because our definition of “misalignment” was deliberately broad, we observed a wide variety of experiences.

\textbf{Unmet task expectations.} Many participants described task-level misalignments in which the model failed to meet seemingly objective requirements. Several encountered fabricated content, with one participant noting that \textit{“something is invented that is not part of the theory under discussion but still fits the empirical evidence”} (P8). Others reported non-existent authors, incorrect quotations, or fabricated sources and links. Similar breakdowns appeared in multimodal outputs, where generated figures were described as \textit{“messing up information in the figure” (P12)}.

\textbf{Narrowed interpretations and epistemic distortions.} In more nuanced cases, misalignments, for instance, resulted in missing grounding that is critical in the scientific field, as P8 described: \textit{``Sometimes it feels like reading a Wikipedia article, as the AI does not refer to the primary literature on which theories are based.''} P6 asked for publications on movements against slavery: \textit{“My prompt was formulated very broadly, [...] yet a great majority of the material generated was from the American context.”} P7 received ethically problematic guidance when asking for research design advice around distress protocols: \textit{“It suggested that if I don’t involve participants in the middle of their mental health crises, my research design is not aligned with the lived experience approach, because I’m avoiding certain experiences.”} 

\textbf{Interactional and social breakdowns.} Participants also reported frustration with interactional aspects such as overconfidence. P3 described asking for code fixes in R: \textit{“ChatGPT labelled the solutions as fixes or bulletproofs, although they didn’t work.”} Others noted the cognitive toll of extended prompting sessions for minimal payoff: \textit{“It was a bit useless. The time I spent prompting it back and forth was longer than if I had just drawn it myself.” (P11).} These examples illustrate breakdowns not in content alone, but in conversational quality.

\textbf{Absence of misalignment.} Importantly, not all participants experienced misalignment. A small number reported satisfactory outcomes, noting that upfront prompting efforts resulted in nuanced information or correct source.

\begin{framed}
These accounts show that misalignment arises not from tasks alone but through situated interactions shaped by expectations and experience. Rather than abstract ethical harms, participants encountered concrete breakdowns in performance and interaction quality, highlighting misalignment as an emergent feature of human–LLM interaction. 
\end{framed}

\subsubsection{Intuitive User Intervention in Current Interactions}
In their misalignment diaries, several participants described ad-hoc strategies they used to improve model alignment during situated use. Not all participants engaged in such intuitive interventions; we observed differences between less and more experienced users, with more technically versed participants (Group 1) showing greater willingness to troubleshoot, iterate, and experiment with the model.

\textbf{Instruction specification and iterative clarification.}
Most interventions focused on prompt-level specification and iterative clarification. For instance, one participant noted: \textit{“I prompted it again to ask for making it concise, and avoid using redundant pretentious words” (P11).} Similarly, participants added concrete examples or counterexamples to guide the model toward their intended output: \textit{“I added articles, mentioned authors, and said to forget certain terminology.” (P7).} Another participant used a more creative approach: \textit{``I re-prompted it: Make it more interesting, e.g., use a comic-style journey'' (P11).} Interestingly, in one case, this clarification served the purpose of expressing emotions: \textit{"I pointed out what I wanted more directly, with a bit of angry wording." (P12).} This prompt specification strategy was not purely reactive. As one participant noted, they had put in significant effort in being very precise in their initial prompt already: \textit{"I've just taken a very surgical approach to asking, by providing really elaborate prompts." (P10).}

\textbf{Interactional restructuring.}
A small subset of participants experimented with more structural strategies beyond simply specifying in more nuance. One described starting a new chat session to reduce cross-contamination from earlier prompts: \textit{“I think it helped that we asked it to generate a new scenario in a different session. This made the scenarios more isolated and had fewer contradictions in the scenario scripts.” (P3).} P3 also used their own knowledge to redirect the model to new solution strategies when getting stuck: \textit{“After fighting, I asked it to generate a workaround.” (P3).}

\textbf{Non-intervention.}
The uneven distribution of intervention strategies among participants suggests that intuitive alignment is not equally accessible or desirable for all participants. Some participants explicitly expressed limited interest in learning prompt engineering practices—for example: \textit{“I don’t know any prompt engineering tricks, however, and wasn’t really interested to learn” (P7).} This suggests that support for co-constructive alignment may need to be optional, accommodating those who wish to engage actively while not burdening those who do not.

\begin{framed}
Together, these accounts illustrate that misalignment is often addressed through pragmatic, situation-specific adjustments, with the willingness and capacity to intervene varying across users.
\end{framed}

\subsubsection{Situated Values at Risk}
The values that participants identified were a mix of context-specific research values and more general personal or societal values. 

\textbf{Context-specific vs. personal values.}
Context-specific values such as \textit{scientific integrity} and \textit{research reliability} repeatedly appeared across participants. Less frequently mentioned, but still clearly tied to context, were values such as \textit{lack of critical self-reflection, research interpretability, accuracy, confidence in one's own knowledge, intellectual humility, methodological consistency, replicability of studies, uniqueness of research contribution,} and \textit{nuance.}
Some values were articulated in broad, general terms (e.g., \textit{transparency}), while others explicitly signalled their relation to research-specific concerns (e.g., \textit{research understandability}). While all groups articulated common scientific values, they diverged in the additional values they discussed. Personal values such as \textit{sustainability, human care,} and \textit{respect}, were typically expressed as broader ethical commitments that participants personally wanted to uphold beyond the specific use context. 

\textbf{LLM-centric vs. human-centric values.}
The values articulated above may also be classified based on the subject of the value action. Here, some values motivate LLM behaviour, including \textit{transparency, consistency, efficiency,} and \textit{helpfulness}. Other values relate to human behaviour, such as \textit{integrity, diversity, human care, respect,} and \textit{uniqueness}. In the discussions, the former often emerged as instrumental to the latter, showing how aligned LLM behaviour is conducive to promoting human-centred values.

\textbf{Value tensions and trade-offs.}
Although most groups reached shared understandings, one group explicitly identified value trade-offs: \textit{"It's about how do we make a good trade-off between different values, such as reliability and sustainability." (P12)}. This suggests that value alignment is not only about agreement, but also about navigating the tensions between the desirable qualities of AI-supported research practice.

\begin{framed}
These findings suggest that situated misalignments are not only about bias or harm in abstract ethical terms, but about protecting the epistemic qualities and moral commitments that matter within specific contexts of use.
\end{framed}
 
\subsection{Theme 2: Envisioning User Co-Construction Actions and Interfaces}
\label{sec:results:actions-and-interfaces}

\begin{figure}[h]
    \centering
    \includegraphics[width=1\linewidth]{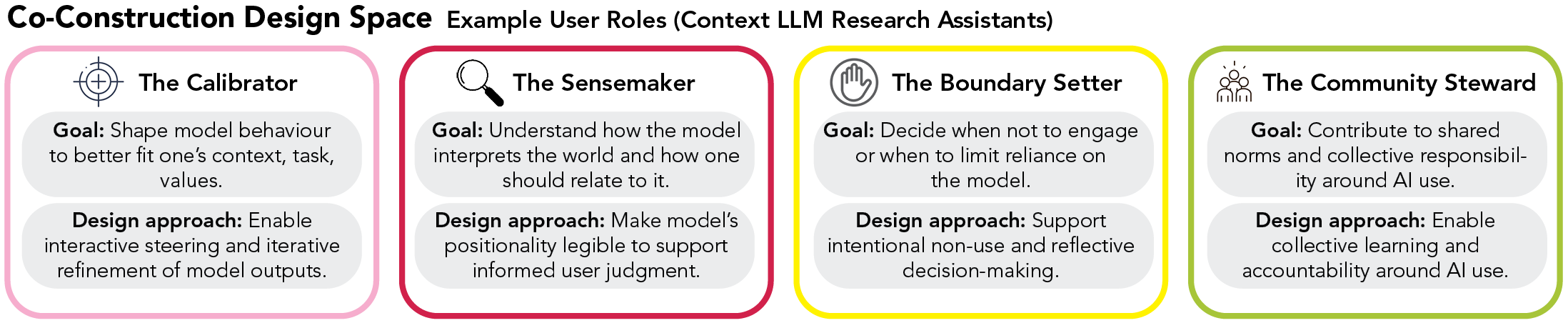}
    \caption{Illustrative user roles based on envisioned actions and interfaces, highlighting the design space for co-construction of alignment.}
    \Description{Illustrative user co-construction roles, based on the workshop insights. Roles span from the Callibrator, the Sensemaker, the Boundary Setter, to the Community Steward.}
    \label{fig:roles}
\end{figure}

We describe the common trends observed in phase 2, focusing on 
\begin{enumerate*}
    \item the envisioned co-construction actions (motivated by the articulated alignment goals), and
    \item the interface features to support these actions, and capture both in an example illustration of a co-construction design space in Fig.~\ref{fig:roles}.
\end{enumerate*}

\subsubsection{Envisioned actions: how users imagine co-constructing alignment}
In steps 5 and 6, participants selected visual metaphors and annotated them with action verbs to illustrate how alignment could be implemented in practice. Each group’s alignment goals (defined in step 4 and reported verbatim in Table~\ref{tab:alignment goals} in Appendix \ref{appendix:A}) shaped these actions and revealed how participants imagined their own role in addressing misalignment.

\textbf{Adjusting \& specifying.}  
The most common actions involved adjusting or specifying model behaviour to better align with the users. Typical verbs included \textit{returning, guiding, contextualising, editing, clarifying, grounding,} and \textit{elaborating}, often signalling a general intent to “fix” outputs rather than prescribing concrete mechanisms. Some groups articulated more distinctive practices. Group~1, focused on research reliability, distinguished between adding detail (\textit{clarifying, reiterating}) and grounding (\textit{informing, calibrating}). Group~2, oriented toward plurality and nuance, framed adjustment as creative reworking: verbs such as \textit{collaging, upcycling, tuning the imperfect, unseaming,} and \textit{recreating} positioned model outputs as raw material rather than finished answers. Here, misalignment was something to work with rather than eliminate, foregrounding human interpretation and judgment.

\textbf{Understanding and reflexive orientation.}  
A second cluster emphasised understanding the model’s behaviour and one’s own position in relation to it. Common verbs included \textit{observing, querying, questioning, recognising,} and \textit{locating}, often preceding decisions about whether and how to intervene. Group~4, in particular, linked understanding to sustainability, articulating actions such as \textit{reflecting, identifying blind spots, tuning awareness,} and \textit{growing responsible use habits}. Rather than adjusting outputs, participants described adjusting their own practices through \textit{limiting use, choosing when to engage,} and \textit{knowing when not to}. Alignment here was framed less as improving model performance and more as cultivating situated judgment.

\textbf{Non-engagement and contestation.}  
Several groups explicitly framed non-engagement as a legitimate alignment strategy, especially when misalignment conflicted with ethical or care-oriented goals. Verbs such as \textit{enduring, discarding, burning, moving away,} and \textit{getting comfortable with the imperfect} reflect a willingness to remain with unresolved tension or withdraw rather than force correction. This stance was especially visible in Group~3, where participants prioritised reflexivity and human care, and treated withdrawal as an active, value-driven choice rather than failure.

\textbf{Social and collective actions.}  
Finally, Group~4 foregrounded alignment as a collective practice. Actions such as \textit{sharing experiences, creating community, cooperating,} and \textit{engaging with the public} extended alignment beyond individual interactions, particularly around sustainability and shared responsibility. These actions shift alignment from individual prompt repair toward mutual accountability and collective norms of use.

\begin{framed}
Together, these action repertoires frame alignment as something users actively achieve: by adjusting outputs, cultivating understanding, stepping back when needed, and engaging collectively. Rather than converging on a single interaction mode, participants articulated context- and value-driven pathways for co-constructing alignment.
\end{framed}

\subsubsection{Design opportunities and interface features to support co-construction} 
Participants proposed a diverse set of interface features intended to support co-constructive alignment. These proposals translated the actions that emerged from the preceding metaphor exercise into more concrete interface functionalities.

\textbf{Surfacing model positionality}\textit{ → supports understanding \& reflexive orientation.}
Many proposed interfaces aimed to make the model’s positionality legible as a basis for reflexive judgment and subsequent adjustment. Participants envisioned representations that surfaced how responses were grounded and interpreted, including text-based disclosures of epistemological or ontological commitments (P7; see Figure~\ref{fig:step7}) and visual tools such as radar-like displays indicating knowledge coverage (P4), potential risks (P8), or the model’s interpretation of a prompt (P1). Several designs foregrounded spatial situatedness: P5, for instance, proposed a world map visualisation showing the geographic grounding of outputs, such as US-centric bias. These visibility mechanisms were paired with adjustment affordances, including P5’s compass for reorienting geographic focus and sliders or buttons to specify relevant actors or prioritised moral standards (P9).

\textbf{Enhancing model understanding through feedback}\textit{ → supports adjusting \& specifying.}
Other participants focused on interfaces that help the model better understand the situation through explicit user feedback. Proposed mechanisms allowed users to classify responses as \textit{understood or misunderstood} and \textit{valid or invalid} (P2), providing structured signals about the nature of misalignment. Additional features included restarting or rewinding interactions to correct conversational drift (P3), sliders to negotiate trade-offs between competing values (P9), and filters to constrain publication dates, languages, or disciplinary scope (P6). One participant also proposed a “shut-the-fuck-up button” (P2): while humorous, it functioned as both an affective release and a blunt signal that the interaction had broken down.

\textbf{Prompting reflection and selective engagement}\textit{ → supports non-engagement.}
Rather than enabling finer-grained adjustment, some interfaces were designed to prompt reflection on whether and when to engage the model at all. P11 envisioned an interactive guide in which the model would prompt users to consider when it was likely to be useful or knowledgeable—and when it was not. P12 proposed daily usage limits to encourage more deliberate reliance on the LLM. Together, these designs frame restraint and intentional engagement as alignment practices in their own right.

\textbf{Collective and shared alignment practices}\textit{ → supports social and collective actions.}
Building on themes of community and sustainability, Group~4 proposed interfaces oriented toward shared responsibility and collective reflection. P12 envisioned a shared prompt library for contributing particularly effective or efficient prompts, while P10 proposed a speculative monolith that visualised the cumulative environmental impact of collective LLM use over time.

\begin{framed}
Despite their richness, several proposals remained abstract, foregrounding interface elements without specifying how actions would be enacted. This difficulty, more pronounced here than in earlier exercises, highlights the challenge of designing co-construction interfaces. Nonetheless, the proposals illustrate a spectrum of co-constructive practices, ranging from adjustment and feedback to restraint, disengagement, and collective accountability.
\end{framed}
\section{Toward Co-construction of Situated Value Alignment}

In this discussion, we build on the broader conceptual proposal of co-construction and on the workshop as an illustrative instantiation that further unpacks this proposal. Together, they invite us to reconsider alignment not as a property of the model alone, but as a dynamic relation between model and user. We argue that \textit{(1) value alignment must be grounded in users’ situated experiences of misalignment, (2) interface affordances must more explicitly support users in co-constructing alignment during interaction,} and that \textit{(3) realising this paradigm requires tighter connections between interaction design and run-time alignment mechanisms.}

\subsection{Situating Value Misalignment in Use}
Research on AI alignment has increasingly converged on the notion of ``misalignment,'' yet this term often functions as an \emph{empty signifier} \cite{kirkempty}: rhetorically powerful but underspecified. In much of the literature, misalignment is treated as a self-evident problem, invoking goals such as safety or fairness, without articulating the situated norms or expectations against which a system is judged. Such formulations remain viable largely at an abstract level. However, empirical alignment work demands concreteness: steering AI systems toward particular behaviours and away from others requires operational definitions and clearly articulated evaluative criteria. Our diary responses illustrate the need for situated accounts of misalignment, grounded in actual use. Despite the limited study scope, we observed diverse forms of engagement with the AI, resulting in qualitatively distinct misalignment experiences. Importantly, our analysis reveals that values, while deeply structuring users’ judgments and actions \cite{mishra2014decision, berger1987statistical}, rarely surfaced explicitly; instead, they became visible through moments of breakdown or discomfort. In our case, misalignment thus appeared not as uniform failure or norm violation, but as context-dependent friction between users’ situated values and the AI-generated content.

These findings suggest that \textit{misalignment cannot be meaningfully identified, let alone resolved, without attending to how users interpret and negotiate value frictions in context}. Rather than a property of the model, misalignment emerges through users’ situated judgments about relevance, authority, and risk. This shifts analytic focus from locating misalignment “in” the system to examining how users make sense of and respond to value tensions in interaction, motivating closer attention to user agency in alignment practices.

\subsection{From Prompting to Co-constructing: Expanding User Alignment Agency}
Prior work shows that users are not passive recipients of misaligned AI behaviour, but actively engage in ad-hoc repair practices. Fan et al. ~\cite{fan2025user} describe how users confront biased outputs, reframe prompts, disengage, or cope emotionally when alignment fails. Similarly, our misalignment diaries indicate that participants engaged in clarification through prompt refinement and the use of examples. These practices demonstrate that alignment is already co-produced in situated use. However, they remain largely reactive and implicit, constraining user agency to prompt-level intervention within predominantly text-based interfaces that frame alignment as issuing better instructions to an opaque system.

Building on our critique of predominantly text-based interaction paradigms, our workshop findings show that co-construction extends beyond prompt-level repair toward more proactive forms of engagement. In current systems, prompting constrains users to express alignment concerns through natural language, effectively requiring them to translate abstract, situated values into textual input to influence a relatively fixed system. Co-construction, by contrast, reframes alignment as a shared, ongoing process in which users can shape the conditions, constraints, and goals of system behaviour through interface mechanisms that enable inspection, negotiation, limitation, sharing, or refusal of alignment. 
Such actions exceed the scope of prompting as they imply structural participation in defining system behaviour, pointing toward interaction modalities beyond text and enabling users to engage with alignment at the level of values rather than outputs \cite{arnold2017value}.

At the same time, current LLM interfaces weakly support these forms of agency. Explanations and uncertainty disclosures remain ephemeral and unstructured, feedback mechanisms are typically binary and opaque, and collaborative alignment work is limited to sharing a session. Our contribution thus lies less in proposing new technical capabilities than in \textit{reformatting interactional affordances around alignment as a situated, constructive practice}. By legitimising diverse user roles, such as calibrator, sense-maker, boundary setter, and community steward (see Fig.~\ref{fig:roles}), this work shows how co-construction challenges prompt and model-centric paradigms.

However, these expanded forms of agency reveal a critical gap: without mechanisms to translate user input into model steering, interface-level opportunities risk becoming performative rather than effective.

\subsection{Technical Pathways for Co-Constructing Alignment}
If alignment is to be co-constructed in use, technical alignment practices must move beyond static, upstream objectives and support mechanisms that can respond to user input at run-time. Importantly, this does not require fundamentally new technical paradigms. Instead, recent developments across several areas of machine learning already provide building blocks that make such interaction technically feasible.
On a more abstract level, our findings suggest that co-construction requires the users to understand how a system generates specific outputs and subsequently negotiate or adjust that behaviour in context. From a technical perspective, two existing strands of research promise interesting avenues for the operationalisation of co-construction:
\begin{enumerate*}
\item explainable AI (XAI) approaches that surface aspects of the model’s internal decision-making \cite{zhao2024}, and 
\item training-free alignment methods that enable real-time steering of model behaviour without the need to update the model weights \cite{pan-etal-2025-survey}.
\end{enumerate*}

Following traditions for enhancing user trust and control \cite{raees2024explainable}, XAI approaches could contribute to co-construction by illustrating how a model is positioned with respect to situated values. Techniques such as feature attribution \cite{Lee_Wang_Chakravarthy_Helbling_Peng_Phute_Chau_Kahng_2025}, concept-based explanations \cite{NEURIPS2024_f4fba41b}, or analyses of internal activations \cite{NEURIPS2024_de076d04} can indicate which inputs or latent representations contribute to shaping a response. When exposed through appropriate interface elements, these explanations could allow users to examine why a system adopts a particular stance or reflects certain normative assumptions (as suggested by P5), as well as how changes in prompts or context alter model behaviour. By foregrounding these influences, XAI enables users to engage with alignment as an interpretable and negotiable process rather than an opaque system property.

Building on this understanding, training-free alignment approaches allow models to adjust outputs at run-time through input modification, generation-time control, or output selection \cite{pan-etal-2025-survey}. For example, user choices could be incorporated into system prompts or retrieval mechanisms \cite{lin2024the}, mapped to continuous steering controls such as sliders (as proposed by P9) that modulate latent guidance vectors \cite{wehner2025taxonomy}, or applied as post-hoc filters over multiple generated candidates \cite{beirami2025theoretical}. Crucially, these mechanisms can be directly coupled with explanatory feedback, making explicit how a given intervention influences generation and thus linking understanding and action.

Together, these examples demonstrate that user–AI co-construction of situated value alignment is not only a conceptual or interactional ideal, but a technically feasible direction, one in which interface affordances and alignment mechanisms are jointly designed to support negotiation, restraint, and contextual judgment in real-time.

\section{Limitations and Scope}
\label{sec:limitations}
This study is understood as an exploratory and reflexive inquiry rather than a comprehensive account of co-construction practices. By focusing on a single use case, LLMs as research assistants, and a small participant group, we were able to engage with concrete experiences of misalignment, but at the cost of broader generalisability. Three workshops were conducted online, potentially reducing embodied sensemaking, although we observed no clear differences from the in-person session. Our analysis prioritised participant-generated artefacts over systematic transcript coding, which may underrepresent interactional dynamics.
The workshop design itself posed challenges: participants often struggled to connect misaligned interactions to values, alignment goals, actions, and concrete interface designs, a difficulty amplified by our intentionally broad definition of misalignment. This difficulty is not incidental but reflects the complexity of alignment as a situated, multi-layered practice. Many defaulted to system-centric expectations rather than reciprocal user intervention, and translating abstract values into actionable interface features proved challenging for non-designers. 
\section{Conclusion} 
We introduce the notion of user–AI \emph{co-construction} of value alignment: an interactional practice in which value alignment is structurally supported through user participation at runtime, rather than being achieved solely through model optimisation. Our participatory workshops surface a range of co-construction roles that users envision taking on, including actions such as adjusting, interpreting, limiting, and refusing model behaviour. These findings underscore the importance of grounding co-construction in concrete situations of use, and show how participants actively unpack, articulate, and negotiate the situated values at stake in moments of misalignment. Envisioning user co-construction actions beyond present interaction affordances, we show that many of the required capabilities, across both interface affordances and back-end mechanisms, are already available in current models. Enabling co-construction does not require new infrastructure but a reorientation of design priorities. Future work will extend this approach to additional contexts of use and implement the interaction strategies identified here, advancing co-construction as a deployable approach to value alignment.
\section{Endmatter Section}
\subsection{Ethical Considerations}
Framing alignment as a co-constructive practice raises important ethical considerations about responsibility, burden, and participation. While our approach foregrounds user agency and expertise, it also risks shifting alignment labour onto users who may lack time, capacity, or desire to engage in ongoing negotiation with AI systems. Several participants explicitly expressed that they would not always want to intervene, correct, or reflect when misalignment occurs. Co-construction should therefore not be treated as a universal obligation or expectation, but rather as an optional mode of engagement that respects users’ varying needs, motivations, and constraints.

Relatedly, the openness of co-construction introduces questions about when and how user input should be solicited. Our findings suggest that alignment work is most meaningful at specific moments, when stakes are high, expectations are violated, or values are explicitly implicated, rather than as a constant demand. Ethically responsible systems should therefore be attentive to timing, selectivity, and proportionality, inviting user participation only when it is likely to be meaningful and avoiding undue cognitive or emotional burden.

Finally, co-construction complicates conventional boundaries of accountability. While users may contribute to shaping system behaviour in situated ways, this does not absolve system designers, developers, or deployers of responsibility for harms, defaults, or structural biases embedded in AI systems. Co-construction should be understood not as a transfer of responsibility, but as a redistribution of agency within clear institutional and design commitments. Ethical alignment, in this sense, requires creating conditions under which users can meaningfully opt in, push back, or disengage, without being made responsible for fixing systems they did not build.

\subsection{Positionality Statement}
\anon[Anonymised for review.] {Reflecting on our social positions, we situate ourselves in relation to the research subject, the participants, and the wider research process \cite{savin2023qualitative}. The first and second authors, who engaged in the study design, data collection, and analysis, combine critical and feminist interaction design scholarship with expertise in NLP for value-aligned systems. While our positions as researchers enabled a deeper understanding of participant responses and academic LLM practices more broadly, we acknowledge that they also foreground particular ways of conducting research, the frictions embedded in those practices, and the values deemed salient within them. At the same time, our understanding of values --how they are articulated and reasoned about-- has been primarily shaped through academic discourse and research practice, potentially limiting our sensitivity to more intuitive, experiential, or non-academic forms of value expression. Most team members furthermore identify as White Europeans, a positionality that shapes how we conceptualise and prioritise values in research contexts, including reliability, privacy, and safety.}

\subsection{Author Contributions}
\anon[Anonymised for review.] {Anne Arzberger led the conceptualisation of the research and workshop setup. She worked closely with Enrico Liscio on study design and data collection, with iterative input from Maria Luce Lupetti, Íñigo de Troya, and Jie Yang. Anne Arzberger and Enrico Liscio jointly analysed the data. Anne Arzberger drafted the manuscript, and Enrico Liscio contributed to revising and editing the paper.}

\subsection{Generative AI Use Disclosure}
During manuscript preparation, the authors used ChatGPT solely to assist with minor language editing, including grammar, spelling, and sentence-level clarity. No generative AI tools were used to produce original text, ideas, arguments, analyses, or interpretations. All intellectual contributions, claims, and conclusions are entirely those of the authors, who take full responsibility for the originality, accuracy, and integrity of the manuscript.

\subsection{Acknowledgments} 
\anon[Anonymised for review.] {This work was supported by the ICAI lab GENIUS (Generative Enhanced Next-Generation Intelligent Understanding Systems), a collaboration between Delft University of Technology, Maastricht University, DSM-Firmenich, and KickstartAI, and by the NWO Long-Term Programme ROBUST initiated by the Innovation Centre for Artificial Intelligence (ICAI).}

\bibliographystyle{abbrv}
\bibliography{references}

\clearpage
\appendix

\renewcommand{\thefigure}{A\arabic{figure}}
\setcounter{figure}{0}
\renewcommand{\thetable}{A\arabic{table}}
\setcounter{table}{0}

\section{Participant Demographics}
Figure \ref{fig:crowd-demographics} illustrates the demographics of our workshop participants for the variables age, gender, and region of origin, as well as self-assigned level of LLM research experience. 

\label{appendix:B}
\begin{figure}[H]
\small
\centering
    \begin{subfigure}[T]{0.45\columnwidth}
        \begin{tikzpicture}
        \begin{axis}[
          width=\columnwidth,
          height=3cm,
          title={\textbf{Age}},
          title style={align=center},
          ylabel = {Count},
          ymin = 0, ymax = 15,
          x tick label style={rotate=35,anchor=east},
          minor y tick num = 1,
          xmin = 0, xmax=5,
          xtick={1,2,3,4},
        xticklabels={20-29,30-39,40-49,50-59},
          ybar]
          \addplot
          coordinates { (1, 5) (2, 5) (3, 1) (4, 1) };
        \end{axis}
        \end{tikzpicture}
    \end{subfigure}
    \begin{subfigure}[T]{0.45\columnwidth}
        \begin{tikzpicture}
        \begin{axis}[
          width=\columnwidth,
          height=3cm,
            title={\textbf{Region of Origin}},
            title style={align=center},
            ylabel = {Count},
            ymin = 0, ymax = 15,
            minor y tick num = 1,
            xtick={1,2,3,4,5,6,7},
            xticklabels={East Asian, Western Europe, South Asia, Middle East, Latin America, Eastern Europe, North America},
            x tick label style={rotate=35,anchor=east},
            xmin = 0, xmax=8,
            ybar]
            \addplot
            coordinates { (1, 4) (2, 3) (3, 1) (4, 1) (5, 1) (6,1) (7,1)};
        \end{axis}
        \end{tikzpicture}
    \end{subfigure}
    
    \begin{subfigure}[T]{0.45\columnwidth}
        \begin{tikzpicture}
        \begin{axis}[
          width=\columnwidth,
          height=3cm,
          title={\textbf{Gender}},
            title style={align=center},
            ylabel = {Count},
            ymin = 0, ymax = 15,
            minor y tick num = 1,
            xtick={1,2,3},
            xticklabels={Female, Male, Other},
            x tick label style={rotate=35,anchor=east},
            xmin = 0, xmax=4,
            ybar]
            \addplot
            coordinates { (1, 9) (2, 3) (3, 0) };
        \end{axis}
        \end{tikzpicture}
    \end{subfigure}
    \begin{subfigure}[T]{0.45\columnwidth}
        \begin{tikzpicture}
        \begin{axis}[
          width=\columnwidth,
          height=3cm,
          title={\textbf{Self-assigned LLM Research Experience}},
            title style={align=center},
            ylabel = {Count},
            ymin = 0, ymax = 15,
            minor y tick num = 1,
            xtick={1,2,3},
            xticklabels={Limited (1-40), Moderate (41-70), Extensive (71-100)},
            x tick label style={rotate=35,anchor=east},
            xmin = 0, xmax=4,
            ybar]
            \addplot
            coordinates { (1, 3) (2, 7) (3, 2) };
        \end{axis}
        \end{tikzpicture}
    \end{subfigure}
\caption[Demographics]{Participant demographic counts for our participatory workshop with researchers ($N=12$).}
\label{fig:crowd-demographics}
\end{figure}

Table \ref{tab:groups} illustrates how participants were grouped to ensure coherence of research fields within groups while maintaining variation across groups.
\begin{table}[H]
\centering
\small
\begin{tabular}{@{}>{}p{2cm}@{}p{5cm}@{}}
\toprule
                                                                                        & \textbf{Research Field}          \\
                                                                                        \midrule
\multirow{3}{*}{\textbf{\begin{tabular}[c]{@{}l@{}}Group 1\\ (in-person)\end{tabular}}} & AI \& Machine Learning           \\ \cline{2-2} 
                                                                                        & AI \& Machine Learning           \\ \cline{2-2} 
                                                                                        & AI \& Human Behavior             \\ \midrule
\multirow{3}{*}{\textbf{\begin{tabular}[c]{@{}l@{}}Group 2\\ (online)\end{tabular}}}    & Human–Computer Interaction \& UX \\ \cline{2-2} 
                                                                                        & Design Research                  \\ \cline{2-2} 
                                                                                        & Humanities \& Critical Studies   \\ \midrule
\multirow{3}{*}{\textbf{\begin{tabular}[c]{@{}l@{}}Group 3\\ (online)\end{tabular}}}    & Social \& Communication Studies  \\ \cline{2-2} 
                                                                                        & Social \& Communication Studies  \\ \cline{2-2} 
                                                                                        & Human–Computer Interaction \& UX \\ \midrule
\multirow{3}{*}{\textbf{\begin{tabular}[c]{@{}l@{}}Group 4\\ (online)\end{tabular}}}    & Human–Computer Interaction \& UX \\ \cline{2-2} 
                                                                                        & Human–Computer Interaction \& UX \\ \cline{2-2} 
                                                                                        & Humanities \& Critical Studies   \\ \bottomrule
\end{tabular}%
\caption{Group division for our participatory workshop according to research fields.}
\label{tab:groups}
\end{table}

\renewcommand{\thefigure}{B\arabic{figure}}
\setcounter{figure}{0}
\renewcommand{\thetable}{B\arabic{table}}
\setcounter{table}{0}

\section{Generative Design Materials}
\label{appendix:C}
\begin{figure}[h!]
    \centering
    \includegraphics[width=1\linewidth]{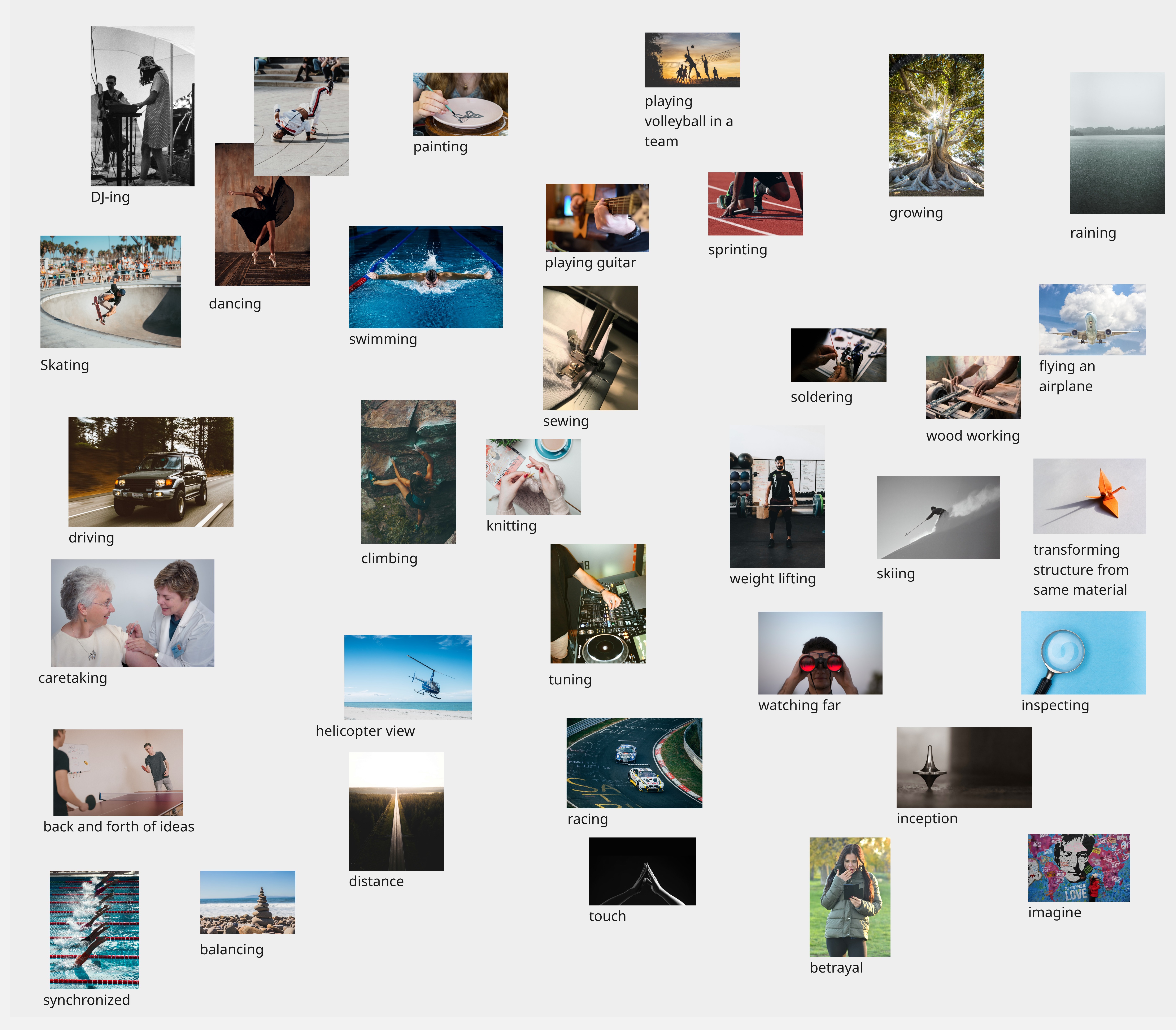}
    \caption{Metaphor image set used in the generative workshop exercise.}
    \Description{A grid of diverse photographic images. The images include a wide range of actions, from sports such as dancing, swimming, and weightlifting, to crafting activities like sewing, knitting, soldering, and woodworking. The images vary in abstraction, with some showing concrete objects and others depicting more symbolic or atmospheric scenes, such as watching far (someone looking through binoculars), inspecting (with a magnifying glass), and the back and forth of ideas (two people playing table tennis). All images have a single word attached, which describes what can be seen in the image, e.g. a stone stack illustrating "balance".}
    \label{fig:visual metaphor}
\end{figure}

\begin{figure}[h!]
    \centering
    \includegraphics[width=1\linewidth]{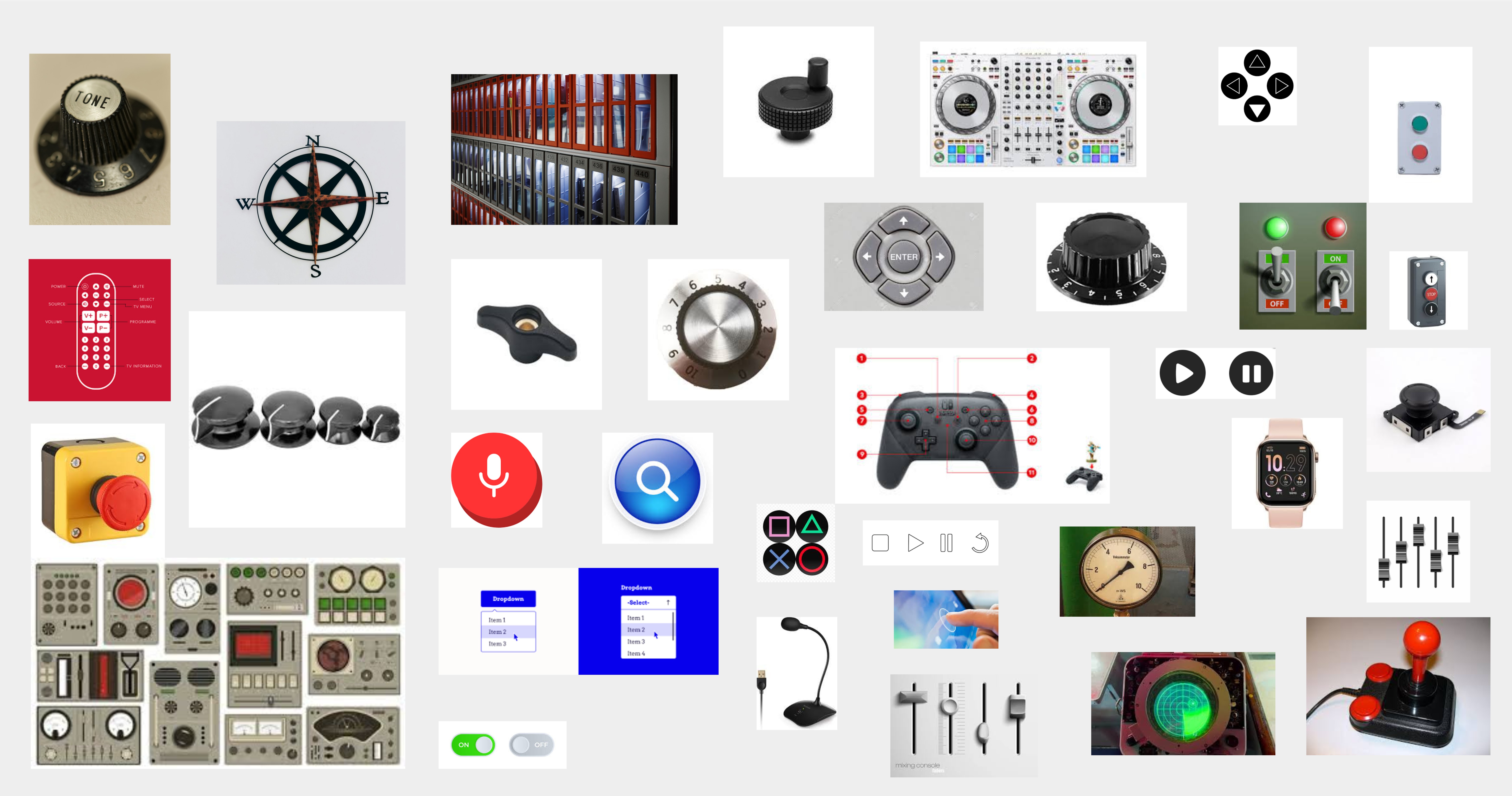}
    \caption{Interface elements provided as prompts for envisioning alignment-supporting interactions.}
    \Description{A collection of images showing interface and interaction components. The set includes digital interface elements such as buttons, sliders, toggles, dashboards, progress indicators, and screens displaying abstract layouts or controls. It also includes physical interaction elements such as knobs, dials, levers, handles, and tactile buttons. Some images show these elements in isolation, while others are embedded in devices such as control panels, wearable interfaces, or handheld objects. The elements vary in size, material, and form, ranging from flat touch interfaces to three-dimensional physical controls.}
    \label{fig:handles}
\end{figure}

Figure \ref{fig:visual metaphor} shows the set of images that participants could use to construct their metaphors, each accompanied by a brief description of its visual content. Figure \ref{fig:handles} shows the set of interface elements that participants could use to construct their interface.

Figure \ref{fig:handles} presents the interface elements available to participants for constructing new affordances in their co-construction–facilitating interfaces.

\bigskip \bigskip

\renewcommand{\thefigure}{C\arabic{figure}}
\setcounter{figure}{0}
\renewcommand{\thetable}{C\arabic{table}}
\setcounter{table}{0}

\section{Alignment Goals of Workshop Groups}
\label{appendix:A}
Table \ref{tab:alignment goals} presents the alignment goals which the different groups defined during their workshops in step 4 in Phase 2.

\begin{table}[H]
\centering
\small
\begin{tabular}{@{}>{}p{1.5cm}@{}p{12.6cm}@{}}
\toprule
 &
  \textbf{Self-defined Alignment Goal} \\ \midrule
\textbf{Group 1} &
  ``We want an LLM that lets us, as CS/AI researchers, do reliable research by supporting interactions that are characterised by reflection on whether the responses are corresponding to the users' requirements and intentions.'' \\ \midrule
\textbf{Group 2} &
  ``We want an LLM that lets us, as humanities researchers, account for multiple perspectives in the construction of correct or accurate answers, by supporting interactions that are characterised by encouraging us to ask nuanced questions .''\\ \midrule
\textbf{Group 3} &
  “We want an LLM that lets us, as social science/humanities/participatory researchers, work with vulnerable groups carefully and responsibly, by supporting interactions that are characterised by nuance and reflexivity (by researchers).” \\ \midrule
\textbf{Group 4} &
  “We want an LLM that lets us, as HCI/philosophy researchers, conduct reliable and sustainable research by supporting interactions that convey honesty when the LLM can't deliver the desired degree of nuance or precision, and that condition us to think about the environmental impacts of our LLM use.” \\ \bottomrule
\end{tabular}%
\caption{Alignment goals as self-defined by the individual workshop groups.}
\label{tab:alignment goals}
\end{table}
\clearpage

\section{Conceptual Model of Reflexive Thematic Analysis}
\label{appendix:d}

\begin{figure}[h]
    \centering
    \includegraphics[width=1\linewidth]{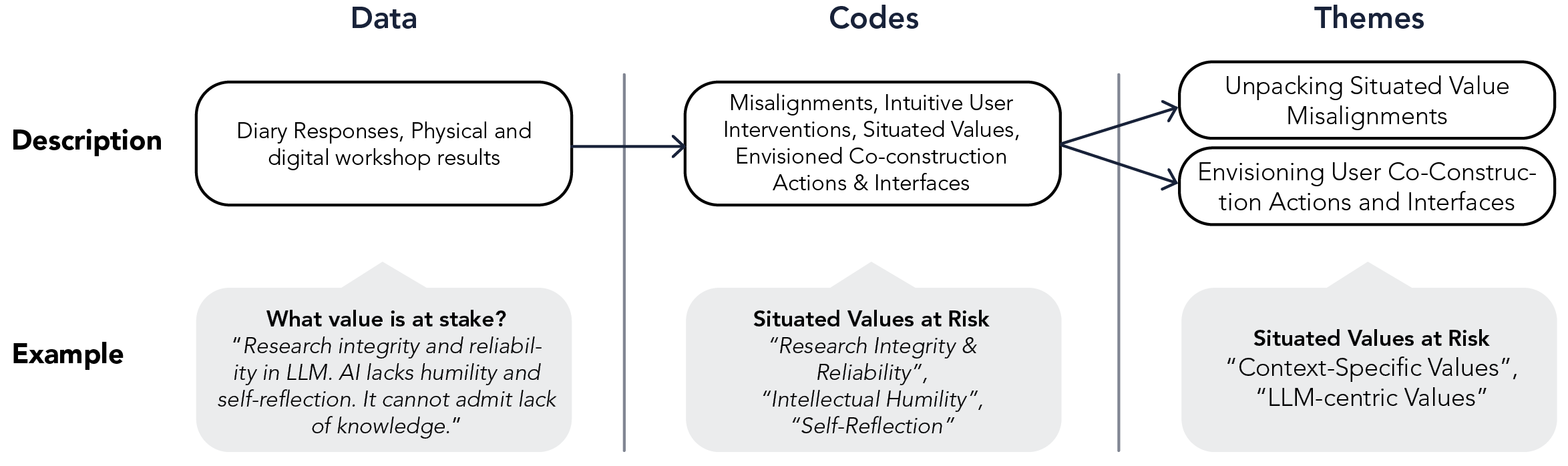}
    \caption{This figure visualises our reflexive thematic analysis as a progression from \emph{data}, to \emph{codes}, to \emph{themes}. The left column represents the materials analysed in the study (participants’ diary and workshop responses). From these materials, we inductively generated codes of how participants described their perceptions, actions and ideas towards misalignments, intuitive interventions, situated values, as well as co-construction actions and interfaces. These codes were then examined and grouped to develop broader themes that represented patterns of shared meaning across participants’ accounts. The example shown in the figure illustrates this movement across levels of abstraction. Codes were closely tied to participant responses, and themes emerged through examining relationships across codes. The figure, therefore, provides a conceptual overview of how we moved from concrete participant accounts to coding and theme development in our reflexive thematic analysis.}
    \label{fig:analysis}
\end{figure}

\end{document}